\documentclass[a4paper]{spie}  
\usepackage[dvipdfm]{graphicx}
\usepackage{ulem}
\usepackage{color}

\title{A Warm Near-Infrared High-Resolution Spectrograph with
Very High Throughput (WINERED)}

\author{Sohei~Kondo\supit{a}, Yuji~Ikeda\supit{b,a}, Naoto~Kobayashi\supit{c,d,a}, Chikako~Yasui\supit{e,a}, Hiroyuki~Mito\supit{d,a}, Kei~Fukue\supit{e,a},
Kenshi~Nakanishi\supit{f}, Takafumi~Kawanishi\supit{g,a},
Tetsuya~Nakaoka\supit{g,a}, Shogo~Otsubo\supit{g,a} Masaomi~Kinoshita\supit{h,g}, Ayaka~Kitano\supit{g,a},
Satoshi~Hamano\supit{c}, Misaki~Mizumoto\supit{e,i},
Ryo~Yamamoto\supit{c}, Natsuko~Izumi\supit{c},
Noriyuki~Matsunaga\supit{e,a}, Hideyo~Kawakita\supit{a,g}
\skiplinehalf 
\supit{a}Laboratory of Infrared High-resolution
spectroscopy (LIH), Koyama Astronomical Observatory, Kyoto-Sangyo
University, Motoyama, Kamigamo, Kita-ku, Kyoto, 603-8555, Japan;\\
\supit{b}Photocoding, 460-102 Iwakura-Naka-machi, Sakyo-Ku, Kyoto
606-0025, Japan;\\ 
\supit{c}Institute of Astronomy, School of Science, the University of
Tokyo, 2-21-1 Osawa, Mitaka, Tokyo, 181-0015, Japan;\\ 
\supit{d}Kiso Observatory, Institute of Astronomy, School of Science, the University of Tokyo, 10762-30
Mitake, Kiso-machi, Kiso-gun, Nagano, 397-0101, Japan;\\
\supit{e}Department of Astronomy, Graduate School of Science, the
University of Tokyo, 7-3-1 Hongo, Bunkyo-ku, Tokyo, 113-0033, Japan;\\
\supit{f}Koyama Astronomical Observatory, Kyoto-Sangyo University,
Kamigamo, Motoyama, Kita-ku, Kyoto, 603-8047, Japan;\\ 
\supit{g}Faculty of Science, Kyoto-Sangyo University, Kamigamo, Motoyama, Kita-ku, Kyoto,
603-8047, Japan;\\ 
\supit{h}Solar-Terrestrial Environment Laboratory, Nagoya University, Furo-cho, Chikusa-ku, Nagoya, Aichi, 464-8601, Japan;\\
\supit{i}Institute of Space and Astronautical
Science (ISAS), Japan Aerospace Exploration Agency (JAXA), 3-1-1,
Yoshinodai, Chuo-ku, Sagamihara, Kanagawa, 252-5210, Japan;}

\authorinfo{Further author information: (Send correspondence to S.K.)\\
S.K.: E-mail: kondosh@cc.kyoto-su.ac.jp, Telephone: +81 75 705 3001\\
Y.I.: E-mail: ikeda@photocoding.com, Telephone: +81 75 708 6120\\
N.K.: E-mail: naoto@ioa.s.u-tokyo.ac.jp, Telephone: +81 422 34 5032}

\begin{document} 
\maketitle 

\begin{abstract}

WINERED is a newly built high-efficiency (throughput$ > 25-30\%$) and
high-resolution spectrograph customized for short NIR bands at 0.9-1.35
${\rm \mu}$m. WINERED is equipped with ambient temperature optics and a
cryogenic camera using a 1.7 ${\rm \mu}$m cut-off HgCdTe HAWAII-2RG
array detector. WINERED has two grating modes: one with a conventional
reflective echelle grating (R$\sim$28,300), which covers
0.9-1.35\,$\mu$m simultaneously, the other with ZnSe or ZnS immersion
grating (R$\sim$100,000). We have completed the development of WINERED
except for the immersion grating, and started engineering and science
observations at the Nasmyth platform of the 1.3~m~Araki Telescope at
Koyama Astronomical Observatory of Kyoto-Sangyo University in Japan. We
confirmed that the spectral resolution ($R\sim$ 28,300) and the
throughput ($>$ 40\% w/o telescope/atmosphere/array QE) meet our
specifications. We measured ambient thermal backgrounds (e.g., 0.06
 ${\rm [e^{-}/sec/pixel]}$ at 287 K), which are
roughly consistent with that we expected. WINERED is a portable
instrument that can be installed at any telescope with Nasmyth focus as
a PI-type instrument. If WINERED is installed on a 10 meter telescope,
the limiting magnitude is expected to be J=18-19, which can provide
high-resolution spectra with high quality even for faint distant
objects.

\end{abstract}



\section{INTRODUCTION}

High resolution spectrograph in near-infrared (NIR) wavelength range is
a powerful tool to explore a variety of astronomical objects from
planets to cosmological objects by measuring chemical abundance and gas
dynamics with atomic and/or molecular lines. We are developing a NIR
high-resolution spectrograph WINERED (Warm Near infrared Echelle
spectrograph to Realize Extreme Dispersion;
Ikeda~et~al. 2006\cite{Ikeda+2006}, Yasui~et~al. 2006\cite{Yasui+2006},
2008\cite{Yasui+2008}). The primary objective of WINERED is to realize
{\it NIR high-resolution spectrograph with high sensitivity} by
achieving high throughput ($>25-30\%$), which is about twice as high as
those of conventional high resolution spectrographs.  WINERED has a wide
wavelength coverage mode, ``Wide-Mode'' with a normal reflective echelle grating, which can
simultaneously cover a wide wavelength range (0.9-1.35\,${\rm \mu m}$)
with a resolving power that is comparable to those of many IR high
resolution spectrographs (R$\sim$29,000;
Yasui~et~al. 2006\cite{Yasui+2006}).  WINERED also aims for the highest
spectral resolution (R$\sim$100,000) by developing ZnSe or ZnS immersion
grating (``High-Resolution-Mode'') while this immersion grating is now under
development\cite{Ikeda+2009,Ikeda+2010}.

Because the wavelength range of WINERED is limited to 0.9-1.35 ${\rm \mu m}$,
where the ambient thermal background is very small, a warm optical
system with no cold stop can be realized. Because of the compact
design (the size is 1.8m $\times$ 1.0m $\times$ 0.5m and the total weight
is $\sim$250 kg), WINERED, which is now located at the Nasmyth
platform of 1.3 m Araki Telescope at Koyama Astronomical
Observatory of Kyoto-Sangyo University, can be moved to any larger
telescopes as a PI-type instrument.

This paper is structured as follows: \S2 shows the optical performance
of WINERED from the engineering observations. \S3 briefly presents our
science grade array and its cassette. \S4 shows the results of the ambient background
measurement. \S5 presents the detection limits of WINERED.  In \S6, we
comment on our future plan.

\section{Optical performances}
\subsection{Overview}

\begin{figure}[!hbt]
\begin{center}
\includegraphics[scale=0.28]{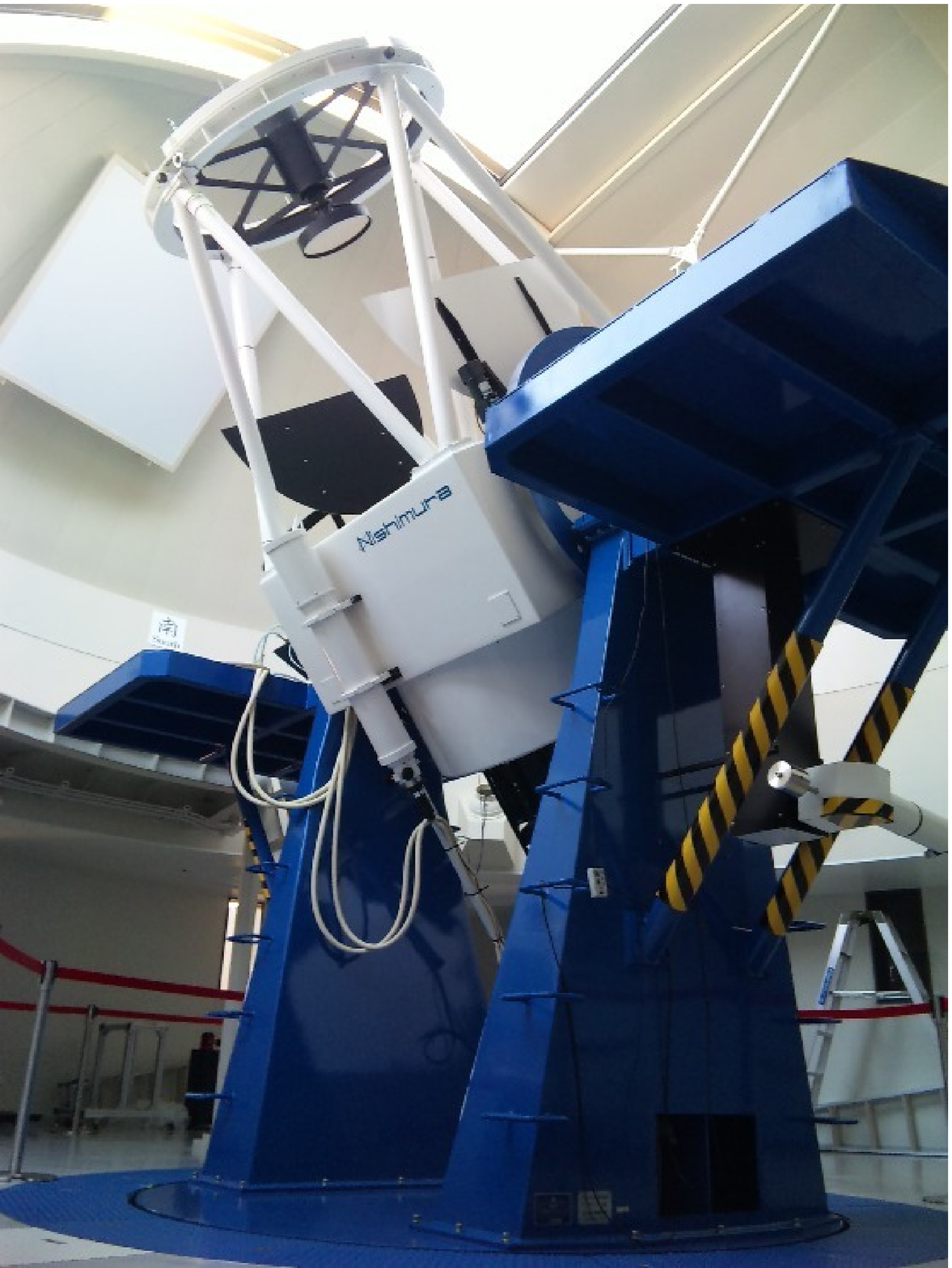}
\includegraphics[scale=0.8]{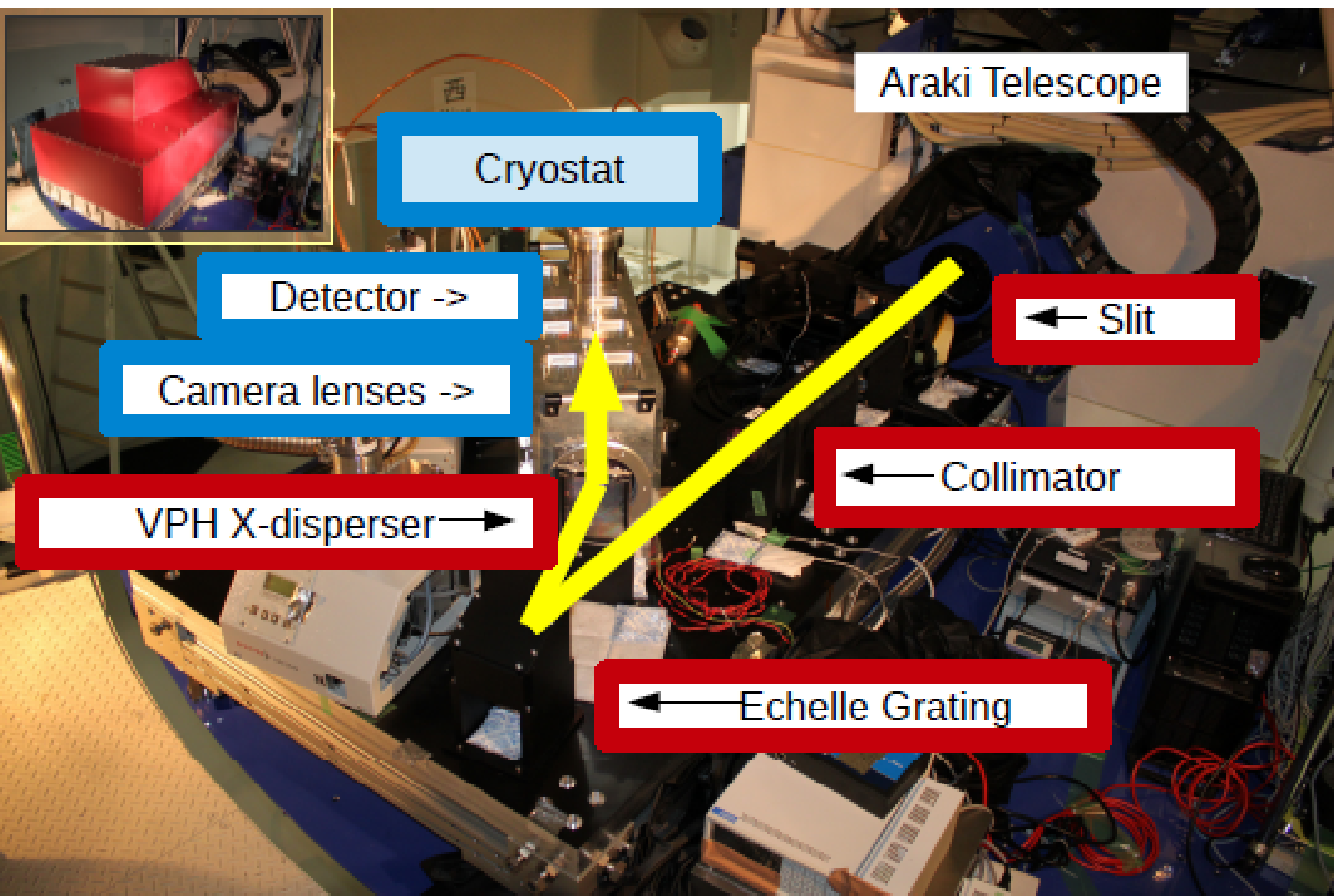} \caption{The 1.3\,m
Araki Telescope at KAO (left), and WINERED installed on the Nasmyth
platform of the telescope (right). The cover of WINERED is removed for
viewing purpose. The slit, the collimator, the echelle grating, and the cross-disperser
(VPH) are in the ambient environment with room temperature. The figure
 in left top corner of this panel shows covered WINERED.\label{fig:winered_araki}} 
\end{center} 
\end{figure}

 WINERED has two observational modes, one is wide wavelength coverage mode
 (``Wide-Mode'') covering 0.90-1.35\,${\rm \mu m}$ in one exposure with
 R=28,300 using a reflective echelle grating. The other is
 high-resolution mode (``High-Resolution-Mode''), which has two setting, $Y$
 and $J$ that cover 0.96-1.13\,${\rm \mu m}$ and 1.12-1.35\,${\rm \mu
 m}$, respectively, with R=103,000 using ZnSe or ZnS immersion
 grating. The optical configuration of WINERED is shown in Figure 2 of
 Yasui et al. (2008)\cite{Yasui+2008}. Overall specifications and
 optical parameters are summarized in Tables \ref{tab:spec} and
 \ref{tab:optical_para}, respectively. At present, WINERED has been
 completed except for the immersion grating. We mount WINERED on the
 Nasmyth focus of the 1.3\,m Araki Telescope of Koyama Astronomical
 Observatory (KAO) at Kyoto-Sangyo University in Kyoto Japan and has
 started engineering and science observations with Wide-Mode (Figure
 \ref{fig:winered_araki}). Almost all optical components are in the
 ambient environment with room temperature except for the camera lenses
 and the infrared array, which are operated with $\sim$90 K and 70 K in
 a cryostat, respectively.

\begin{table}[!ht]
\begin{center}
\small 
\begin{tabular} {ccc}
\hline
\hline
&High-Resolution-Mode&Wide-Mode\\ \hline
Maximum spectral resolution & 103,000 (2-pix sampling)&28,300
	 (2-pix sampling)\\
Wavelength coverage&$Y$: 0.96-1.13\,${\rm \mu m}$ & 
	 0.90-1.35\,${\rm \mu m}$\\
&$J$: 1.12-1.35\,${\rm \mu m}$&\\
Volume&\multicolumn{2}{c}{1800 mm(L) $\times$ 1000 mm(W) $\times$ 500 mm(H)}\\
\hline
\end{tabular}
\caption{WINERED basic specifications.\label{tab:spec}}
\end{center}
\end{table}

\begin{table}[!hbt]
\begin{center}
\small 
\begin{tabular} {ccccc}
\hline
\hline
& &High-Resolution-Mode& Wide-Mode\\\hline
Slit &Width& \multicolumn{2}{c}{100, 200, 400 $\mu$m}\\
&Length &\multicolumn{2}{c}{3.12 mm} \\ \hline
Collimator &Focal length& \multicolumn{2}{c}{770 mm}\\
&Clear aperture&\multicolumn{2}{c}{84 mm}\\ \hline
Echelle&Type&ZnSe or (ZnS) immersion grating&classical echelle grating&\\
&Blaze angle& 70 deg.& 63.9 deg.\\
&Groove density & 31.80 gr/mm&31.60 gr/mm\\\hline 
Cross-disperser &Frequency &710 lines/mm ($Y$)& 280 lines/mm\\ 
&&510 lines/mm ($J$)&\\
&Bragg angle&20.8 deg. ($Y$)&9.3 deg.\\
&&17.9 deg. ($J$)&\\ \hline
Camera&Focal length& \multicolumn{2}{c}{266.80 mm}\\ 
&Clear aperture&\multicolumn{2}{c}{128.25 mm}\\
\hline
Detector &Array format& \multicolumn{2}{c}{2k$\times$2k (Teledyne,
	 HAWAII-2RG)}\\
&Pixel size&\multicolumn{2}{c}{${\rm 18\, \mu m \times 18\, \mu m}$}\\
&Cut-off wavelength&\multicolumn{2}{c}{1.76 ${\rm \mu m}$}\\ \hline 
Slit viewer&FOV&\multicolumn{2}{c}{${\rm 4.8'\times 3.5'}$ (w/ the 1.3\,
	 m Araki Telescope)}\\ 
&Wavelength region&\multicolumn{2}{c}{${\rm 0.6-0.9\, \mu m
}$}\\ \hline
Artificial light source&&\multicolumn{2}{c}{Th-Ar (for wavelength calibration)}\\ 
&&\multicolumn{2}{c}{Halogen lamp}\\ 
\hline
\end{tabular}
\caption{Optical parameters of WINERED.\label{tab:optical_para}}
\end{center}
\end{table}

\subsection{Coverage}
\begin{figure}[!h]
\begin{center}
\includegraphics[scale=1.5]{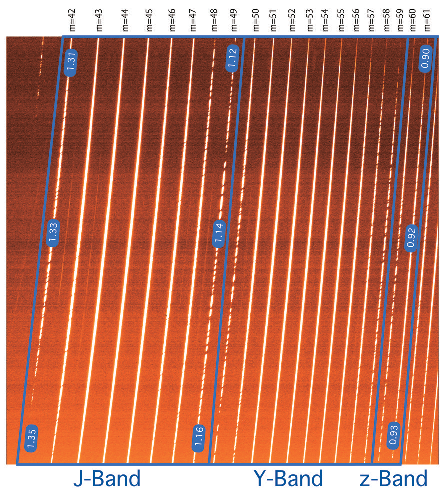}
\includegraphics[scale=1.5]{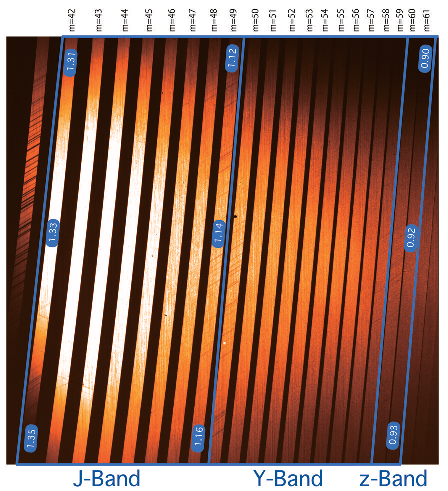} \caption{Echellogram
of $\alpha$ Boo (left) and a flat-lamp (right). The faint
spectra seen between low orders are the ghosts from the 2nd-order
lines of the VPH cross-disperser (HAWAII-2RG has the sensitivity for the
optical wavelength). However, because the ghosts are enough separated
from the object spectrum, they do not produce any critical
problem. \label{fig:echelle_format}}
\end{center}
\end{figure}

\begin{figure}[!ht]
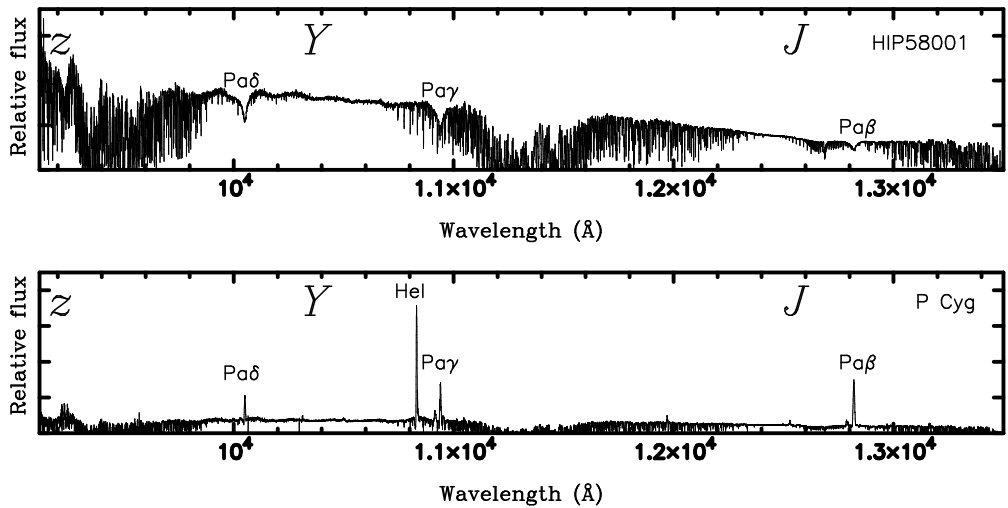

\begin{center}
\includegraphics[scale=0.8]{std_all_v3.eps}\\
\vspace{0.3cm}
\includegraphics[scale=0.8]{pcyg_v3.eps} \caption{Top panel: the
spectrum of a star HIP 58001 (A0V). Broad Pa$\beta,\gamma,\delta$
absorption lines are clearly seen.  The strong telluric absorption
features due to water vapor are seen between $z$, $Y$, and
$J$-bands. Bottom panel: the spectrum of P Cyg. Pa$\beta,\gamma,\delta$,
emission lines as well as very strong HeI emission have clear P Cygni
profiles.  \label{fig:winered_spectrum}}
\end{center}
\end{figure}

Figure \ref{fig:echelle_format} shows the echellograms of $\alpha$ Boo (Arcturus) and a flat-lamp obtained with Wide-Mode. We
confirmed that the entire wavelength range, 0.90-1.35 ${\rm \mu m}$
(m=41-61), is covered in a single exposure by investigating the
echellogram of Th-Ar comparison lamp. Figure \ref{fig:winered_spectrum}
shows the Wide-Mode spectra of a A0V star (HIP 58001) and P Cyg, which
show broad hydrogen absorption lines and strong emission lines,
respectively. This wide wavelength range of about 4,500 {\r A} can be
covered in one exposure, which should enables extensive classifications
of a variety of astronomical objects.

\clearpage

\subsection{Spectral Resolution}
\begin{figure}[!hbt]
\begin{center}
\includegraphics[scale=1.3]{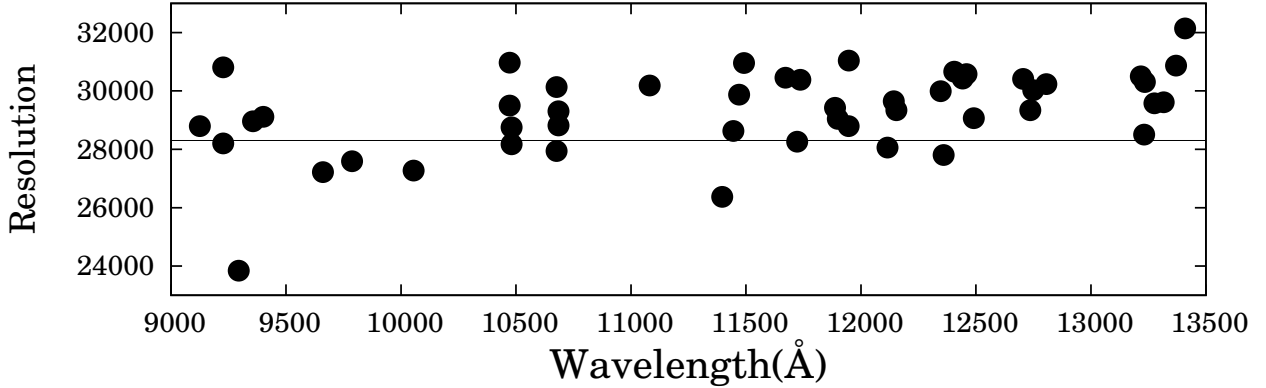} \caption{Measured
spectral resolution for N-mode. The black points show the measured
values. The solid line shows the target spectral resolution, which is
defined by 2-pixel sampling.\label{fig:resolution}}
\end{center}
\end{figure}

We measured the spectral resolution of Wide-Mode using the Th-Ar lamp. The
measured spectral resolutions are defined as the FWHM of single Th-Ar
emission lines. Figure \ref{fig:resolution} shows the obtained spectral
resolution as a function of wavelength. We confirmed that the designed
spectral resolving power ($R=28,300$) is achieved through the
entire wavelength range.

\subsection{Throughput}

\begin{figure}[!hbt]
\begin{center}
\includegraphics[scale=1]{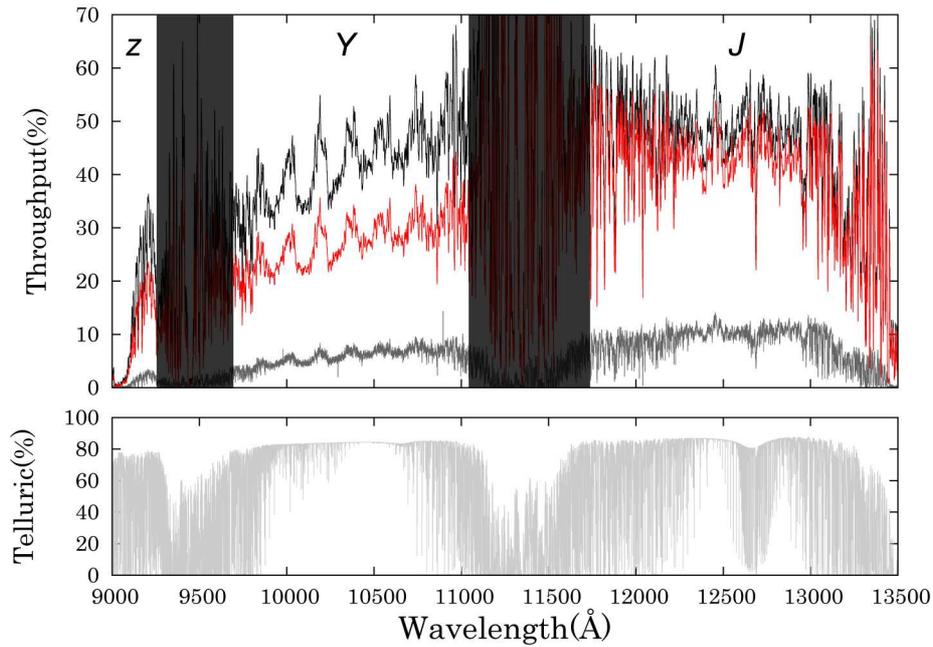}
\caption{Estimated throughput of WINERED for Wide-Mode using EG array. The
top panel shows the throughput (Black: WINERED optics only, Red: WINERED
optics times QE of the SG array, Dark gray: as observed with the EG
array, whose QE is 30-60\% from Teledyne lnc.. The bottom panel shows the assumed telluric
absorption spectrum for estimating the
throughput.\label{fig:winered_throughput}}
\end{center}
\end{figure}

In order to estimate the throughput of WINERED, we observed a photometric
standard star (HD87822), which is listed in the IRTF Spectral
Library\cite{Rayner+2009}, with the 400~${\rm \mu m}$
(=6$^{\prime\prime}$.6) wide slit to avoid the flux loss at the slit
during engineering observation using our engineering array. We assumed
that the efficiency of the telescope, determined by reflectance of
mirrors and vignetting by the baffle for the secondary mirror and the
pupil aperture (This is because WINERED is designed for f/11 telescopes
though f-number of Araki telescope is 10), is about 0.5 from the past
measurements of the telescope. The atmospheric absorption at the KAO
site is calculated with LBLRTM code (Clough et
al. 2005\cite{Clough+2005}) accessing the HITRAN database (Figure
\ref{fig:winered_throughput}: bottom panel). The obtained throughput of
the optics as a function of wavelength is shown in Figure
\ref{fig:winered_throughput}. While the black curve in the figure is in
the case without the EG array, the
red curve is in the case with the SG array assumed. The throughput included an
array QE in $J$-band is found to be over 40\% as designed. However, the
throughput at shorter wavelengths is unexpectedly degraded (down to 20\%
at z-band). We consider that the aerosol scattering is more efficient in
the actual city environment than we expected in our calculation, but
more investigation is necessary.

\section{Infrared array}

We use a 1.7 ${\rm \mu m}$ cut-off 2k$\times$2k HAWAII-2RG
array\cite{Beletic+2008} to suppress ambient thermal backgrounds at
longer wavelengths beyond H-band, and SIDECAR ASIC and
JADE2\cite{Loose+2007} for readout electronics. A science grade (SG)
array has been installed.

\subsection{Array Cassette}

\begin{figure}[!h]
\begin{center}
\includegraphics[scale=0.4]{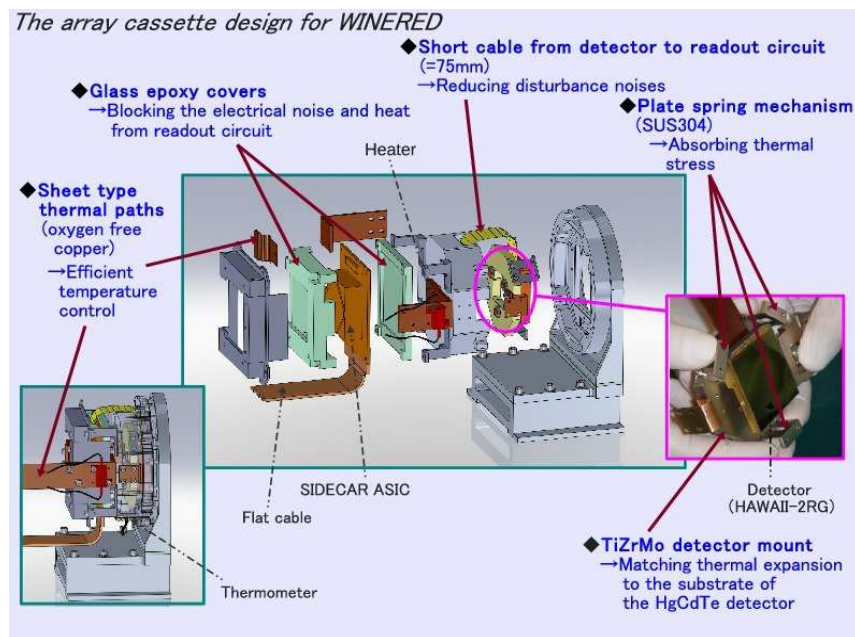}
\caption{Array cassette.\label{fig:winered_fig_mecha}}
\end{center}
\end{figure}

Figure \ref{fig:winered_fig_mecha} shows the new design of our array
cassette. We designed this cassette for safe assembly, releasing thermal 
stress, and easily cooling to the purpose temperature.

\subsection{Performance of the SG array}


The performances of the SG array are summarized in Table
\ref{tab:detector}. The quantum efficiency (QE) was measured by
Teledyne lnc.. Readout noise was measured from the variance of dark frames
with short integration time (15 sec) for which Poisson noise from the
dark electrons is negligible. With the Fowler-sampling, the readout
noise decreases from 19.2$\pm$ 2.9 $[{\rm e^{-}}]$ (NDR=1) to 5.3$\pm$
1.0 $[{\rm e^{-}}]$ (NDR=32). The dark current was estimated from the
ramp sampling over 1,500 sec and is found to be ${\rm 7.6\pm 0.2\times
10^{-3}\, [e^{-}/s]}$. Conclusively, we can say that this SG array meets
our specifications.

The conversion gain is set to be 2.27 ${\rm e^{-}/ADU}$ for the detector
bias of 0.25 v. Readout time is about 1.45 sec per frame for 32-ch
output operation mode with 100 kHz pixel rate. The detector is reset 4
times before readout, so it takes at least 10 sec to obtain one frame
even for the minimum integration time. The counts of the output frame
are corrected with those of the reference pixels. To reduce readout
noise, we use Fowler-sampling of 2, 4, 8, 16 non-destructive reads
depending on the integration time during actual observations.

\begin{table}[h]
\begin{center}
\small 
\begin{tabular} {cccccc}
\hline
\hline
QE [\%]&  Readout noise (NDR=1) [${\rm e^-}$]& Readout noise (NDR=32) [${\rm e^-}$]
 &Dark [${\rm e^-/s}$]& Full well [${\rm e^-}$]&   \\ \hline
63-114 & 19.2$\pm$2.9& 5.3$\pm$1.0& $7.6\pm 0.2 \times 10^{-3}$& $1.4\times 10^5$&  \\
 \hline
\end{tabular}
\caption{Science grade array performance. QE is provided
 by Teledyne lnc.. The uncertainty of QE is probably over
 10\% (from Teledyne lnc.). \label{tab:detector}}
\end{center}
\end{table}


\section{Ambient thermal background}

All optical components except for the camera lens and the infrared array
are placed under the ambient temperature. To block the ambient thermal
background over 1.35\,${\rm \mu m}$ as much as possible, a thermal cut
filter is coated on the cold camera lens in front of the array (Yasui et
al. 2008\cite{Yasui+2008}), and additional thermal blockers (PK50 and
a custom thermal cut filter) are installed. When the ambient temperature
is sufficiently low, the noise from the ambient thermal background is expected to be less
than the readout noise $({\rm \sim 5\,e^-})$ by combination of the
thermal cut filter, the thermal blockers and a 1.7\,${\rm \mu m}$
cut-off array.

We measured the ambient thermal background by putting a black cover on the
window of the cryostat so that the detector looks at a black body with
the room temperature. We confirmed that leak of light is negligible for
this measurement. A cold mask with two holes at the center/edge was
installed at the 4 mm distance from the array. The hole size is 3.2 mm
which is determined as no vignetting for the full FOV of the camera
lens. We measured the ambient thermal background in the bright region and
estimated the dark current and detector bias in the shadow region caused
by the mask simultaneously, which were fount to be negligible.

\begin{figure}[!htb]
\begin{center}
\includegraphics[scale=0.8]{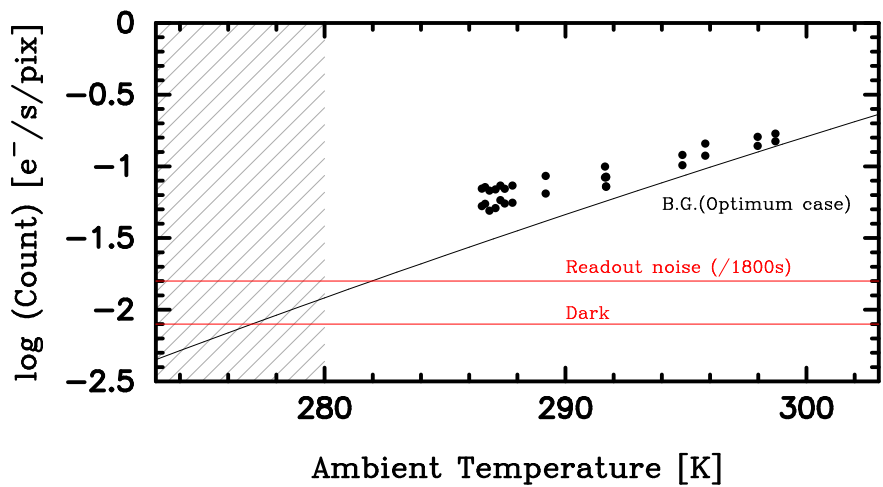}
\caption{Measured ambient thermal backgrounds. The black points are the
measured values. The black line is the expected ambient thermal background in
the optimum case. The red lines show an equivalent readout noise for 1800
sec and the dark current. Hatched region shows the nominal operating
temperature of WINERED.  \label{fig:temp-count}}
\end{center}
\end{figure}

We measured the ambient thermal background only at the lab temperatures
(287--299 K), which is higher than the typical operation
temperatures we expect on the telescopes. The relation between the
ambient temperature and photon counts is shown in Figure
\ref{fig:temp-count}. This figure shows that the measured ambient
background counts are well correlated with ambient temperatures (${\rm \sim
0.06\, [e^-/sec/pix]}$ at 287 K and ${\rm \sim 0.14\, [e^-/sec/pix]}$ at
299 K) and are
roughly consistent with those we expected for the temperatures. There
are some differences of counts between the holes, but the ratios are
almost constant for all the temperatures. The noise from the ambient thermal
background is expected to be less than readout noise under $\sim 280$ K if the ambient thermal background decreases with decreasing ambient
temperature like optimal case. To further verify this expectation, we
will measure ambient thermal backgrounds at low temperatures ($<$ 280 K) in the
telescope dome during the cold winter months.

\section{Detection Limit}

Table \ref{tab:detection_limit} summarizes the estimated limiting
magnitudes of WINERED for various telescopes. The ambient background
count highly depends on the environment. For Araki Telescope, we adopt
the ambient thermal background at around 290 K, which is about an
average ambient temperature throughout the year. For the other
telescopes, we adopt the ambient thermal background at 273 K. The value is
extrapolated from our measured thermal background at 287--299 K, assuming
that the logarithm of the ambient thermal background decreases linearly with
decreasing temperature. Table \ref{tab:detection_limit} shows that goal
magnitudes are achieved if ambient backgrounds decrease as we expect. If WINERED is installed on a 10 meter telescope, the
limiting magnitude is expected to be ${\rm m_J}$=18-19, which can
provide high-resolution spectra with high quality even for faint
objects.

\begin{table}[!h]
\begin{center}
\small
\begin{tabular} {ccccc}
\hline
\hline
&Araki1.3m&WHT4.2m&Magellan6.5m&Keck10m\\ \hline
Location &KAO&Roque de los Muchachos&Las Campanas&Mauna Kea\\
&Kyoto, Japan&La Palma, Spain&Chile&Hawaii\\ \hline
Seeing& $3^{\prime\prime}.0$&$0^{\prime\prime}.8$&$0^{\prime\prime}.6$&$0^{\prime\prime}.4\ (0^{\prime\prime}.2)$\\ \hline
Pixel scale (${\rm /pix}$)&$0^{\prime\prime}.82$&0$^{\prime\prime}$.23&0$^{\prime\prime}$.15&0$^{\prime\prime}$.098\\
Slit width for $R_{max}$&1$^{\prime\prime}$.65&0$^{\prime\prime}$.47&0$^{\prime\prime}$.30&0$^{\prime\prime}$.20\\ \hline
Ambient Temperature (K)&290&273&273&273\\
Thermal Background (${\rm e^- /s /pixel}$) &0.08&0.01&0.01&0.01\\ \hline
Goal $m_{J}$&12.8&15.9&16.5&17.6 (18.4)\\
$m_{J}$&13.1&16.6&17.4&18.3 (19.1)\\ \hline
\hline
\end{tabular}
\caption{Estimated detection-limit of WINERED of Wide-Mode in $J$-band
  for the total integration time of 8 hrs (1800 sec$\times$16) and
  S/N=30. Goal $m_J$, and $m_J$ are the magnitudes when the parameters
  (e.g., throughput, QE of a detector, and the ambient background) of
  Ikeda et al. (2006)\cite{Ikeda+2006} and of this paper are assumed,
  respectively. For the case with the Keck telescope, the use of a focal
  reducer from f/15 to f/11 is assumed. In the line of Keck,
  the values in parentheses are shown using
  AO.\label{tab:detection_limit}}
\end{center}
\end{table}


\section{Current status and future plan}

Since the first light on May 23 2012, we have conducted engineering and
science observation four times to obtain $\sim$200 spectra of a variety
of astronomical objects. The ZnSe or ZnS immersion grating is being
developed and the detail will be reported elsewhere. We plan to
fabricate the final large immersion grating (probably with ZnSe) and to
install it to complete High-Resolution-Mode of WINERED.

\acknowledgments 

We are grateful to Y. Shinnaka for providing us the atmospheric
absorption spectrum at the KAO site. We would like to thank the staffs
of KAO, Kyoto-Sangyo University.  This study was financially supported by
KAKENHI (16684001) Grant-in-Aid for Young Scientists (A), KAKENHI
(20340042) Grant-in-Aid for Scientific Research (B), KAKENHI (26287028)
Grant-in-Aid for Scientific Research (B), KAKENHI (21840052)
Grant-in-Aid for Young Scientists (Start-up). This study has been
financially supported by the MEXT- Supported Program for the Strategic
Research Foundation at Private Universities, 2008-2012 (No. S0801061)
and 2014-2018 (No. S1411028).  S.H. is supported by Grant-in-Aid for
JSPS Fellows Grant Number 13J10504.


\bibliography{report}   

\begin{thebibliography}{1}  

\bibitem{Ikeda+2006} Y.~Ikeda, N.~Kobayashi, S.~Kondo, C.~Yasui, and
	K.~Motohara, ``WINERED: A warm high-resolution near-infrared
	spectrograph,'' in {\it Ground-based and Airborne Instrumentation for Astronomy. Edited
by McLean, Ian S. Iye, Masanori., Proceedings of the SPIE, Volume 6269,
	pp. 62693T (2006), Presented at the Society of
Photo-Optical Instrumentation Engineers (SPIE) Conference} {\bf 6269}, pp. 62693T, July, 2006

\bibitem{Yasui+2006} C.~Yasui, Y.~Ikeda, N.~Kobayashi, S.~Kondo, and
	K.~Motohara, ``Optical design of WINERED: warm infrared echelle
	spectrograph,'' in {\it Ground-based and Airborne
	Instrumentation for Astronomy. Edited by McLean, Ian
S. Iye, Masanori. Proceedings of the SPIE, Volume 6269, pp. 62694P (2006)., Presented at the Society of
Photo-Optical Instrumentation Engineers (SPIE) Conference} {\bf 6269},
	pp. 62694P, July, 2006

\bibitem{Yasui+2008} C.~Yasui, S.~Kondo, Y.~Ikeda, A.~Minami,
	M.~Motohara, N.~Kobayashi, ``Warm infrared Echelle
spectrograph (WINERED): testing of optical components and performance evaluation of the optical system,”
in {\it Ground-based and Airborne Instrumentation for Astronomy
	II. Edited by McLean, Ian S. Casali, Mark M. Proceedings of the
	SPIE, Volume 7014, pp. 701433-701433-12 (2008),
Presented at the Society of Photo-Optical Instrumentation Engineers
	(SPIE) Conference} {\bf 7014}, pp. 701433, June, 2008

\bibitem{Ikeda+2009} Y.~Ikeda, N.~Kobayashi, S.~Kondo, C.~Yasui,
	P.~J.~Kuzmenko, H.~Tokoro, and H.~Terada, ``Zinc sulfide and
	zinc selenide immersion gratings for astronomical
	high-resolution spectroscopy: evaluation of internal attenuation
	of bulk materials in the short near-infrared region,'' in {\it
	Optical Engineering}, Volume {\bf 48}, Issue 8,
	pp. 084001-084001-9, August, 2009

\bibitem{Ikeda+2010} Y.~Ikeda, N.~Kobayashi, J.~P.~Kuzmenko,
	S.~L.~Little, C.~Yasui, S.~Kondo, H.~Mito, K.~Nakanishi, and
	Y.~Sarugaku, ``Fabrication and current optical performance of a
	large diamond-machined ZnSe immersion grating,'' in {\it Modern Technologies in Space- and Ground-based Telescopes and Instrumentation. Edited by
Atad-Ettedgui, Eli Lemke, Dietrich. Proceedings of the SPIE, Volume
	7739, pp. 77394G (2010), Presented
at the Society of Photo-Optical Instrumentation Engineers (SPIE)
	Conference} {\bf 7739}, pp. 77394G, July, 2010

\bibitem{Beletic+2008} J.~W.~{Beletic}, R.~{Blank}, D.~{Gulbransen},
	D.~{Lee}, M.~{Loose}, M.~E.~C.~{Piquette}, T.~{Sprafke},
	W.~E.~{Tennant}, M.~{Zandian}, J.~{Zino}, ``Teledyne Imaging
	Sensors: infrared imaging technologies for astronomy and civil
	space'', in {\it High Energy, Optical, and Infrared Detectors
	for Astronomy III. Edited by Dorn, David A. Holland, Andrew D.
proceedings of the SPIE, Volume 7021, pp. 70210H-70210H-14, Presented
at Society of Photo-Optical Instrumentation Engineers (SPIE) Conference} {\bf 7021}, pp. 70210H, August, 2008

\bibitem{Loose+2007} M.~{Loose}, J.~{Beletic}, J.~{Garnett}, M.~{Xu},
	``High-performance focal plane arrays based on the HAWAII-2RG/4G
	and the SIDECAR ASIC'', in {\it Focal Plane Arrays for Space
	Telescopes III. Edited by Grycewicz, Thomas J. Marshall, Cheryl
	J.~Warren, Penny G., Proceedings of the SPIE, Volume 6690,
	pp. 66900C (2007), Presented
at Society of Photo-Optical Instrumentation
	Engineers (SPIE) Conference} {\bf 6690}, pp. 66900C, September, 2007

\bibitem{Rayner+2009}  J.~T.~{Rayner}, M.~C.~{Cushing}, W.~D.~{Vacca},
	``The Infrared Telescope Facility (IRTF) Spectral Library: Cool
	Stars,'' in {\it The Astrophysical Journal Supplement}, Volume
	{\bf 185}, Issue 2, pp. 289-432, December, 2009

\bibitem{Clough+2005} S.~A.~Clough, M.~W.~Shephard, E.~J.~Mlawer,
	J.~S.~Delamere, M.~J.~Iacono, K.~Cady-Pereira, S.~Boukabara,
	P.~D.~Brown, ``Atmospheric radiative transfer modeling: a summary of the
AER codes'' in {\it Journal of Quantitative Spectroscopy and Radiative
	Transfer}, Volume {\bf 91}, Issue 2, pp. 233-244, March, 2005


\end{thebibliography}
\bibliographystyle{spiebib}   

\end{document}